\def\papertitle{Count-Density Networks for Modal Plate Parameter Estimation}
\def\paperauthorA{Rodrigo Diaz}
\def\paperauthorB{Pablo Tablas De Paula}
\def\paperauthorC{David Marttila}
\def\paperauthorD{Ilias Ibnyahya}
\def\paperauthorE{Chin-Yun Yu}
\documentclass[twoside,a4paper]{article}
\usepackage{etoolbox}
\usepackage[taskB]{dafx26challenge}
\usepackage{amsmath,amssymb,amsfonts,amsthm}
\usepackage{siunitx}
\usepackage{euscript}
\usepackage[T1]{fontenc}
\usepackage[utf8]{inputenc}
\usepackage{ifpdf}
\usepackage[english]{babel}
\usepackage{caption}
\usepackage{subfig}
\usepackage{booktabs}
\usepackage{color}
\usepackage{graphicx}
\usepackage{url}
\input glyphtounicode
\ninept

\newcounter{numauth}
\newcounter{listcnt}
\newcommand\authcnt[1]{\ifdefined#1 \stepcounter{numauth} \fi}
\newcommand\addauth[1]{
  \ifdefined#1
    \stepcounter{listcnt}
    \ifnum \value{listcnt}<\value{numauth}
      \appto\authorslist{, #1}
    \else
      \appto\authorslist{~and~#1}
    \fi
  \fi
}
\authcnt{\paperauthorB}
\authcnt{\paperauthorC}
\authcnt{\paperauthorD}
\authcnt{\paperauthorE}
\authcnt{\paperauthorF}
\authcnt{\paperauthorG}
\authcnt{\paperauthorH}
\authcnt{\paperauthorI}
\authcnt{\paperauthorJ}
\def\authorslist{\paperauthorA}
\addauth{\paperauthorB}
\addauth{\paperauthorC}
\addauth{\paperauthorD}
\addauth{\paperauthorE}
\addauth{\paperauthorF}
\addauth{\paperauthorG}
\addauth{\paperauthorH}
\addauth{\paperauthorI}
\addauth{\paperauthorJ}

\usepackage{times}
\ifpdf
  \usepackage[pdftex,
    pdftitle={\papertitle},
    pdfauthor={\authorslist},
    pdfsubject={Proceedings of the 29th International Conference on Digital Audio Effects (DAFx26)},
    colorlinks=false,
    bookmarksnumbered,
    pdfstartview=XYZ]{hyperref}
\else
  \usepackage[dvips,
    pdftitle={\papertitle},
    pdfauthor={\authorslist},
    pdfsubject={Proceedings of the 29th International Conference on Digital Audio Effects (DAFx26)},
    colorlinks=false,
    bookmarksnumbered,
    pdfstartview=XYZ]{hyperref}
\fi

\newcommand{\code}[1]{\begingroup\urlstyle{tt}\nolinkurl{#1}\endgroup}
\title{\papertitle}
\affiliation{\paperauthorA\textsuperscript{1,2},
\paperauthorB\textsuperscript{1}, \paperauthorC\textsuperscript{1},\\
\paperauthorD\textsuperscript{1}, and \paperauthorE\textsuperscript{1}}
{\textsuperscript{1}Queen Mary University of London, London, United Kingdom\\
\textsuperscript{2}Fraunhofer HHI, Berlin, Germany}
\begin{document}
\maketitle
\begin{abstract}
We describe two submissions to Task B of the 1st DAFx Parameter Estimation
Challenge, which estimates an unknown number of modal frequency, decay, and gain
triples from a synthetic plate-reverb impulse response. The first method combines
pooled spectral features with time-domain and absolute-scale conditioning in a
real-valued convolutional count-density network, while the second uses a
complex-valued Transformer count-density network. Both methods jointly infer the
modal count and per-mode attributes directly from the IR. On an independently
generated 100-IR comparison set, the two neural estimators achieve lower overall challenge
error than the evaluated classical baselines, with frequency and decay
estimation substantially more accurate than gain estimation.
\end{abstract}

\section{Introduction}
\label{sec:introduction}
Task B of the challenge asks participants to recover modal parameters from
measured impulse responses (IRs) generated by a reference plate
simulator~\cite{Ducceschi:2026:Challenge}. The target is an unordered set of
modal triples
\begin{equation}
  \theta = \{(\omega_m, \sigma_m, b_m)\}_{m=1}^{\widetilde M},
  \label{eq:task-b-set}
\end{equation}
where $\widetilde M$ is the identified number of modes. Both methods use the
measured displacement IR and its stored amplitude scale, with the shared input
preprocessing described in Section~\ref{ssec:task-b-preprocessing}. At
inference, the estimators do not use the plate's analytical model.

\begin{figure*}[t!]
  \centering
  \includegraphics[width=\textwidth]{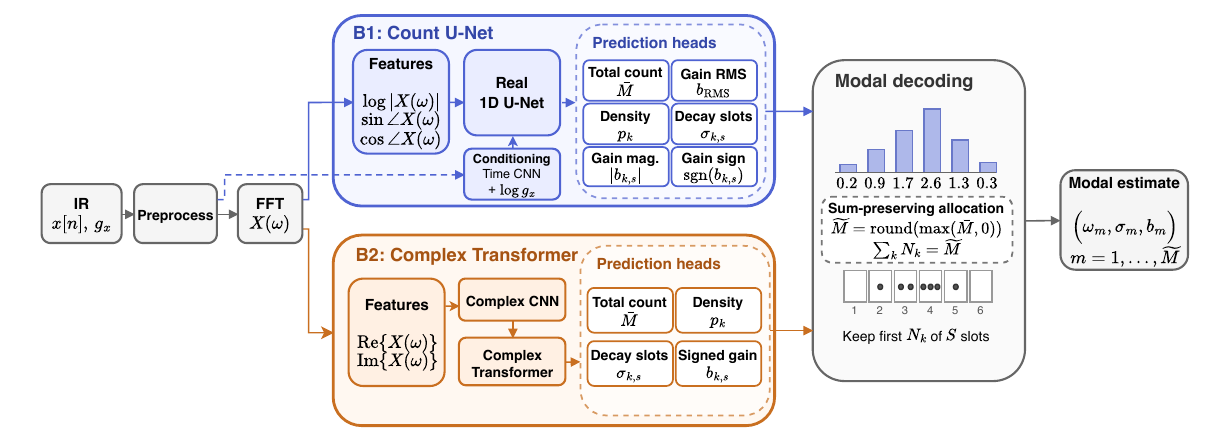}
  \caption{Task B architectures and shared decoding. B1 combines pooled
  magnitude and phase features with a \SI{0.75}{\second} time-domain encoding
  and log-gain conditioning in a real-valued U-Net, while B2 uses the
  full-resolution complex spectrum with a complex CNN and Transformer. The
  sum-preserving allocation combines the predicted total count with the
  modal density, and the resulting per-bin counts select the retained modal
  frequencies and attributes.}
  \label{fig:task-b-overview}
\end{figure*}

\section{Training Data}
\label{sec:training-data}
Training data was generated with a GPU-accelerated implementation of the
reference plate model. With the generator defaults, each of the 20 HDF5 shards
contains $512 \times 32 = 16{,}384$ five-second IRs at
\SI{44.1}{\kilo\hertz}, giving $327{,}680$ IRs in total.

Each row contains the waveform and input-scale fields described in
Section~\ref{ssec:task-b-preprocessing}, together with six normalised physical
parameters and three damping parameters.
Ground-truth modal frequencies, decay rates, and gains are computed on the fly
from these parameters using the simulator's analytical formula rather than
stored as mode tables. This analytical mode list is used solely to label the
synthetic training data.

\section{Methods}
\label{sec:task-b}

\subsection{Input Representation and Preprocessing}
\label{ssec:task-b-preprocessing}
Both methods operate on a peak-normalised waveform
$\tilde{x}[n]=x[n]/g_x$, where $g_x=\max_n\lvert x[n]\rvert$ is the peak
amplitude of the unnormalised displacement IR. During training-data generation,
$\tilde{x}$ and $g_x$ are stored as \code{ir} and \code{global_gain},
respectively. For challenge inference, the same pair is obtained from the
\code{ir} and \code{normalization_factor} fields of each \code{.npz} file. The
accompanying \code{.wav} files contain only $\tilde{x}$ and therefore lack the
amplitude scale required to restore the modal gains.

Before feature extraction, both methods downsample $\tilde{x}$ from
\SI{44.1}{\kilo\hertz} to \SI{22.05}{\kilo\hertz}, since the synthetic IRs
contain no modes above \SI{10}{\kilo\hertz}, and retain the FFT bins up to
\SI{10}{\kilo\hertz}. The gain targets are represented relative to the input
peak, $b_m/g_x$, and the predicted gains are multiplied by $g_x$ when writing
the submission CSV files. Method~B1 additionally conditions its bottleneck on
the normalised logarithm of $g_x$.

\subsection{Targets and Training Objective}
\label{ssec:task-b-training}
Both Task B methods use supervised targets derived from the ground-truth mode
list of each synthetic training IR. The modes are accumulated onto a
$1/20$-octave log-frequency grid from \SI{20}{\hertz} to
\SI{10}{\kilo\hertz}, yielding a per-bin modal count. For each bin, both methods
store the decay rates and gains of its modes in a fixed set of ordered slots.
Method~B1 also learns the global modal-gain scale $b_{\mathrm{RMS}}$, so that
mode count and per-mode attributes are learned jointly.

Because the modal density spans orders of magnitude across frequency, all count
terms are evaluated in $\log(1+\cdot)$ space so that densely populated bins do
not dominate the loss. The count objective $\mathcal{L}_{\mathrm{count}}$ is a
weighted sum of loss terms. Huber losses are used for the fine per-bin,
global total-count, cumulative-count, and multi-resolution errors, with the
multi-resolution counts aggregated by factors $2$, $5$, and $10$.
Cross-entropy aligns the predicted density with the normalised target density.
The cumulative and multi-resolution terms enforce consistency at several
frequency scales rather than bin-by-bin alone. Decay and
gain are trained with slot-wise Huber losses masked to the occupied slots;
Method~B1 factorises the gain into a magnitude, a sign (binary cross-entropy),
and the global RMS scale $b_{\mathrm{RMS}}$, whereas Method~B2 regresses a
single signed gain. The
complete objective is
\begin{equation}
  \mathcal{L} = \mathcal{L}_{\mathrm{count}}
  + \lambda_\sigma\,\mathcal{L}_{\sigma}
  + \lambda_b\,\mathcal{L}_{b},
  \label{eq:task-b-loss}
\end{equation}
where $\mathcal{L}_{\sigma}$ and $\mathcal{L}_{b}$ are the decay and gain terms.
At inference, both methods use the same decoder. The non-negative global count
estimate $\bar M$ is rounded to $\widetilde M$, the per-bin density scores are
rescaled to sum to $\widetilde M$ and floored, and the remaining modes are
assigned to the bins with the largest fractional remainders. Thus,
sum-preserving allocation produces integer bin counts satisfying
$N_k\in\mathbb{Z}_{\geq 0}$ and $\sum_k N_k=\widetilde M$.
Figure~\ref{fig:task-b-overview} summarises the shared inference pipeline and
the two method-specific backbones.

\subsection{Method B1: U-Net Count Density Network}
\label{ssec:task-b-count}
Method B1 uses a neural count-density U-Net~\cite{ronneberger_unet_2015}.
Input features are computed online from the shared peak-normalised IR. The input
has three channels: row-normalised log magnitude, $\sin$ phase, and $\cos$ phase.
Each channel is mean-pooled over FFT bins falling in high-resolution
log-frequency intervals of $1/120$ octave from \SI{20}{\hertz} to
\SI{10}{\kilo\hertz}. No second normalisation is applied after this pooling.

In parallel, a strided time-domain CNN encodes the first \SI{0.75}{\second} of
the decimated IR into a 128-dimensional conditioning vector. This encoding is
concatenated with the normalised logarithm of the stored input scale $g_x$ and
projected into the U-Net bottleneck.

The network has four encoder stages with channel widths 32, 64, 128, and 256,
a 512-channel bottleneck, and four mirrored decoder stages with skip
connections. Each stage uses two one-dimensional convolutional layers with ReLU
activations. Downsampling is max-pooling, and upsampling repeats the decoder
features before concatenating the corresponding skip. The bottleneck feature map
is averaged over frequency and passed to two global heads: one predicts the
total modal-count estimate $\bar M$, and the other predicts
$b_{\mathrm{RMS}}$. Decoder features are
mean-pooled by a factor of six, so the output grid is $1/20$ octave. The model
has approximately 3.1 million parameters.

The decoder-grid heads predict modal-density logits, per-bin decay slots, gain
magnitude slots, and gain-sign logits. The density logits are softmax-normalised
over the $1/20$-octave output bins, giving a distribution over where the modes
should lie in frequency. At inference, the shared count decoder converts this
distribution into integer per-bin counts $N_k$. Frequencies are then placed deterministically inside each
occupied bin. For the attributes, each output bin has a fixed capacity of
$S=768$ ordered modal slots.
If the decoded count in bin $k$ is $N_k$, the export takes the first $N_k$
slots from the decay, gain-magnitude, and gain-sign heads for that bin and
discards the remaining slots. The selected gains are formed from the predicted
sign, magnitude, bottleneck-predicted $b_{\mathrm{RMS}}$, and stored input
scale $g_x$ before
writing the challenge CSV files.

Training used AdamW with a fixed learning rate of $10^{-3}$ and weight decay
$10^{-4}$, a batch size of 64, and 26 epochs, taking roughly 24 hours on a
single NVIDIA A100 GPU. The final checkpoint achieved the lowest validation
Task B relative error after sum-preserving allocation.

\begin{table*}[t!]
  \centering
  \small
  \setlength{\tabcolsep}{3pt}
  \caption{Task B comparison on the independently generated 100-IR comparison
  set. Error cells report mean (variance) across the IRs. Runtime is
  inference-only and excludes I/O. Neural and classical timings use different
  hardware and are therefore not directly comparable. Lower is better.}
  \label{tab:task-b-comparison}
  \begin{tabular}{@{}lcccccc@{}}
    \toprule
    Method & $RE_{\Omega}$ & $RE_{\sigma}$ & $RE_b$ & $\min(1,\Delta M/M)$ & $RE$ & $t_{\mathrm{IR}}$ (s) \\
    \midrule
    B1 (ours)     & \textbf{0.0077 (0.0001)} & 0.0508 (0.0005) & \textbf{0.8525 (0.0001)} & \textbf{0.0143 (0.0002)} & \textbf{0.3180 (0.0002)} & 0.0010 \\
    B2 (ours)     & 0.0080 (0.0002) & \textbf{0.0159 (0.0002)} & 0.9207 (0.0004) & 0.0152 (0.0003) & 0.3301 (0.0006) & 0.0030 \\
    \midrule
    ESPRIT        & 0.6569 (0.0425) & 0.9372 (0.0014) & 0.9649 (0.0004) & 0.6626 (0.0366) & 1.5156 (0.0771) & 442.0 \\
    Matrix Pencil & 0.6566 (0.0425) & 0.9372 (0.0014) & 0.9648 (0.0004) & 0.6622 (0.0368) & 1.5150 (0.0775) & 17.4 \\
    Root-MUSIC    & 0.6727 (0.0397) & 0.9708 (0.0003) & 0.9693 (0.0003) & 0.6768 (0.0349) & 1.5477 (0.0696) & 252.9 \\
    Burg          & 0.2062 (0.0510) & 0.9250 (0.0005) & 0.9138 (0.0006) & 0.4367 (0.0892) & 1.1184 (0.0996) & 210.2 \\
    Fast Burg     & 0.1965 (0.0492) & 0.9038 (0.0007) & 0.9088 (0.0007) & 0.4287 (0.0905) & 1.0984 (0.1017) & 113.5 \\
    Yule--Walker  & 0.1816 (0.0474) & 0.9100 (0.0006) & 0.9071 (0.0007) & 0.4340 (0.0990) & 1.1002 (0.1072) & 147.8 \\
    CWT           & 0.9751 (0.0001) & 0.9775 (0.0001) & 1.0000 (0.0000) & 0.9750 (0.0001) & 1.9592 (0.0003) & 0.7 \\
    \bottomrule
  \end{tabular}
\end{table*}

\subsection{Method B2: Complex-Valued Transformer Count-Density Network}
\label{ssec:task-cv-transformer}
Method B2 uses the same frequency range as Method~B1. It applies
phase-preserving log compression and per-IR RMS normalisation to the
full-resolution complex spectrum, then uses its real and imaginary parts as
features. A five-layer
complex CNN tokenises the spectrum into a sequence of 64 complex tokens. A
six-layer complex-valued Transformer encoder, following Eilers and
Jiang~\cite{Eilers:2023:Building}, uses a model dimension of 320, eight attention
heads, and a feed-forward width of 320. It applies complex attention, complex
layer normalisation, and complex linear maps realised as a pair of real maps
$A+iB$ before a 2048-unit dense bottleneck and the count, decay, and gain heads.
The count objectives otherwise follow Method~B1, but unlike B1's separate
magnitude and sign heads, B2 directly predicts a single signed gain value.
Counting each complex weight as a real/imaginary pair, the backbone totals
approximately 99.6 million real-valued trainable parameters, most of which sit
in the dense bottleneck (83.9 million) rather than the CNN and Transformer
layers themselves (9.4 million combined).

The released checkpoint is from the seventh and final training epoch (epoch 6
under zero-based numbering) and is published on the HuggingFace
Hub.\footnote{\url{https://huggingface.co/ptablasdepaula/complex-valued-modal-plate-transformer}}
Training used AdamW with weight decay $10^{-4}$ and global-norm gradient
clipping (max norm 1.0), a batch size of 32, and seven epochs. The learning rate
warmed linearly from 0 to $10^{-4}$ over 1000 steps, then followed cosine decay
towards $10^{-6}$ over a 100{,}000-step horizon. Training took roughly 27 hours
on a single NVIDIA A100 40~GB GPU. The released checkpoint achieved the lowest
validation Task B relative error after sum-preserving allocation.
\section{Comparison with Classical Methods}
We evaluate every method on the same independently generated comparison set,
produced with the baseline reference-plate generator and its default parameter
ranges. It contains 100 five-second IRs sampled at \SI{44.1}{\kilo\hertz} and
is separate from the HDF5 training shards described in
Section~\ref{sec:training-data}.

Table~\ref{tab:task-b-comparison} compares our two neural estimators with the
classical baselines. Following the challenge notation, the evaluation metric is
\begin{equation}
  \begin{aligned}
    \Delta M &:= \lvert M-\widetilde M\rvert, \\
    RE_0 &= \frac{1}{3}\left(RE_{\Omega}+RE_{\sigma}+RE_b\right)\in[0,1], \\
    RE &:= RE_0+\lambda\min\left(1,\frac{\Delta M}{M}\right),
    \qquad \lambda=1.
  \end{aligned}
  \label{eq:task-b-evaluation}
\end{equation}
Here, $M$ and $\widetilde M$ are the reference and identified in-band mode
counts, respectively, and $RE_0$ averages the frequency, decay-rate, and gain
errors.

We benchmark three band-wise subspace estimators (ESPRIT, Matrix Pencil, and
Root-MUSIC), three autoregressive estimators (Burg, Fast Burg, and
Yule--Walker), and an FFT--CWT peak detector. The parametric methods use 32
equal-width bands, a cap of 100 modes per band, and a within-band decimation
factor of 32; the autoregressive model order is 200. The three autoregressive
methods use three CLEAN passes without band overlap. CWT uses a
\SI{3}{\decibel} prominence threshold, \SI{2}{\hertz} minimum peak spacing,
and a \SI{-40}{\decibel} SNR floor.

All metrics use the same 100 five-second IRs over
\SI{20}{\hertz}--\SI{10}{\kilo\hertz}. Timings exclude input loading and output
writing. Neural timings report batch-one feedforward latency on one A100 GPU,
averaged over 20 passes after three warm-up passes. Classical timings report
the complete estimator computation on 24 cores of an AMD EPYC 9555 CPU,
including band extraction, pole estimation, modal merging, CLEAN processing,
and gain fitting.
The GPU and CPU timings characterise each implementation on its recorded
hardware rather than a controlled cross-hardware speed-up.
The relative cost of these stages varies between estimators. The ESPRIT timing
is approximate because its run required resumption.

\section{Conclusion}
\label{sec:conclusion}
We presented two methods for modal-parameter estimation in Task B. Both methods
use count-density networks that jointly infer an unknown mode count and the
corresponding frequency, decay, and gain triples from the IR. Method~B1 combines
a real-valued U-Net with time-domain and scale conditioning, while Method~B2
directly processes complex values with a complex-valued Transformer. On the
independently generated 100-IR comparison set, B1 and B2 achieve overall relative errors of $0.3180$
and $0.3301$, respectively, compared with $1.0984$ for the strongest evaluated
classical baseline. Gain estimation remains the dominant source of error. The
implementation repository is available online.\footnote{\href{https://github.com/davidmarttila/dafx_challenge}{\nolinkurl{github.com/davidmarttila/dafx_challenge}}}

\bibliographystyle{IEEEtranDAFx}
\bibliography{references}
\end{document}